\documentclass{ws-p8-50x6-00}

\usepackage{graphicx,amsfonts}

\newtheorem{defn}{Definition}
\newtheorem{conj}{Conjecture}

\newcommand{\s}{{}^* \negthinspace}

\newcommand{\M}{{\cal M}}

\newcommand{\R}{\mathbb{R}}

\newcommand{\plus}{\mathop{\!+\!}}

\newcommand{\minus}{\mathop{\!-\!}}

\newcommand{\smallfrac}[2]{\mbox{\small $\frac{#1}{#2}$}}

\renewcommand{\vec}[1]{\textbf{ #1}}

\newcommand{\text}[1]{\mbox{#1}}

\begin{document}

\title{Cosmological Models with Isotropic Singularities}
\author{ Susan M. Scott}
\address{Department of Physics and Theoretical Physics, Faculty of Science,\\
        Australian National University, Canberra, ACT 0200, Australia}
\author{Geoffery Ericksson}
\address{Department of Theoretical Physics, RSPhysSE, IAS,\\
        Australian National University, Canberra, ACT 0200, Australia}

\maketitle                      
\abstracts{
In $1985$ Goode and Wainwright devised the concept of an isotropic
singularity. Since that time, numerous authors have explored the
interesting consequences, in mathematical cosmology, of assuming the
existence of this type of singularity. In this paper, we collate all
examples of cosmological models which are known to admit an isotropic
singularity, and make a number of observations regarding their general
characteristics.
}


\section{Introduction}

In 1985 Goode and Wainwright\cite{GW85} introduced the concept of an
isotropic singularity (IS) to the field of mathematical cosmology, in order to
clarify what is meant by a ``quasi-isotropic'' singularity and a
``Friedmann-like'' singularity. Their definition of an IS is
coordinate-independent, as well as being independent of the Einstein field
equations (EFE), and hence of the source of the gravitational field.
Scott\cite{Scott88a,Scott88b,SSGVR} has amended their definition to remove
some redundancy, and it is this amended definition that will be used
throughout this paper.

\begin{defn}[Isotropic Singularity]
A space-time $(\M ,\vec{g})$ is said to admit an \emph{isotropic singularity}
 if there exists a space-time $(\s\M , \s \vec{g})$, a smooth cosmic time 
function $T$
defined on $\s\M$, and a conformal factor $\Omega (T)$ which satisfy
\begin{enumerate}
\item $\M$ is the open submanifold $T>0$,
\item $\vec{g}= \Omega^{2}(T) \s \vec{g}$ on $\M$, with $\s \vec{g}$ regular
(at least $C^{3}$ and non-degenerate) on an open neighbourhood of $T=0$,
\item $\Omega(0) = 0$ and $\exists\thinspace b>0$ such that
        $\Omega \in C^{0}[0,b] \cap  C^{3}(0,b]$ and $\Omega(0,b] >0$,
\item $\lambda \equiv \lim_{T \rightarrow 0^{+}} L(T)$ exists,
        $\lambda\neq 1$, where
        $L \equiv \frac{\Omega''}{\Omega}
        {\left( \frac{\Omega}{\Omega'} \right)}^{2}$
        and a prime denotes differentiation with respect to T.
\end{enumerate}
\end{defn}

\begin{figure}[!ht]
\centering
\psfig{figure=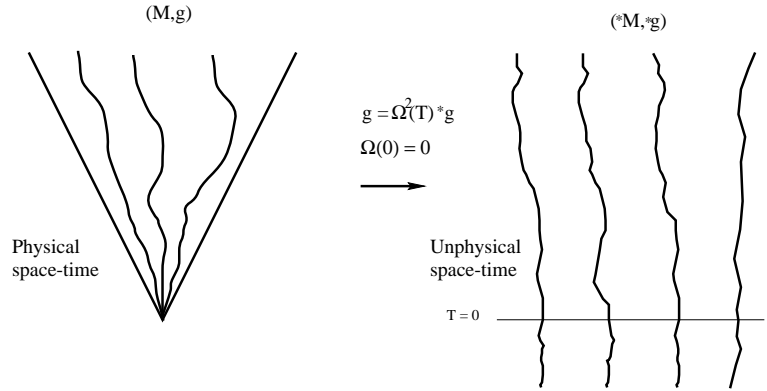}
\caption{Pictorial interpretation of an isotropic singularity.}
\label{is_pic}
\end{figure}

\noindent
The definition of an isotropic singularity is described pictorially in
Figure (\ref{is_pic}).

Since $(\M,\vec{g})$ is the space-time in hand---in our case, typically, a
cosmological solution of the Einstein field equations---it is usual to call it
the \emph{ physical space-time}. The conformally related space-time,
$(\s\M, \s \vec{g})$, is then called the \emph{ unphysical space-time}.
 Although the
definition of an IS, and the arrow in Figure (\ref{is_pic}), lead one to think
of the unphysical space-time being ``created'' from the physical space-time,
it is actually useful to consider the situation in the reverse fashion. In
this way, the singularity in the physical space-time should be regarded as
arising due to the vanishing of the conformal factor, $\Omega(T)$, at the
regular hypersurface $T=0$ in $\s\M$.


The definition of an isotropic singularity, as it stands,  allows cosmological
models to admit an ``isotropic'' singularity whose singularities are, in no
sense, actually isotropic, or quasi-isotropic, or
Friedmann-like.\cite[pp117-8]{Goode82,GW85}  For example, the exact viscous
fluid FRW cosmology of Coley and Tupper\cite{CT83} can be shown to have an IS,
and yet the shear and acceleration of the fluid flow are not dominated by its
expansion as the singularity is approached, as one would expect with a
Friedmann-like singularity. The fluid congruence is not regular at the IS,
however, which motivated Goode and Wainwright\cite{GW85} to include the
following additional definition relating to the fluid flow.

\begin{defn}[Fluid Congruence]
\label{defn:fluid_congruence}
With any time-like congruence $\vec{u}$ in $\M$ we can associate a time-like
congruence $\s \vec{u}$ in $\s \M$ such that
\be
\s \vec{u} = \Omega \vec{u} \quad \text{in}\:\M \, .
\end{equation}

\noindent (a) If we can choose $\s \vec{u}$ to be regular (at least $C^3$)
on an open
neighbourhood of $T=0$ in $\s\M$, we say that $\vec{u}$ is
\emph{ regular at the isotropic singularity}. \\
(b) If, in addition, $\s \vec{u}$ is orthogonal to $T=0$ we say that $\vec{u}$
 is \emph{ orthogonal to the isotropic singularity}.
\end{defn}

\noindent
It is the requirement that the fluid flow be regular at the IS (condition (a)
of Definition \ref{defn:fluid_congruence}), which ensures\cite{GW85} that the
appropriate kinematic quantities behave as one would expect as an
``isotropic'' singularity is approached.

A question which naturally follows from these two definitions is
\emph{ what cosmological models actually admit an isotropic singularity
 at which the fluid flow is regular?}  In Sections $2,3,4,$ and $5$, we
collect together the various examples which have appeared throughout the
literature, and in Section $6$, we discuss their general characteristics,
and draw certain conclusions from them.


\section{Kantowski-Sachs models}
The Kantowski-Sachs models\cite{Wainwright84,Kantowski66} are irrotational,
 geodesic, perfect fluid models with a radiation equation of state
$p=\frac{1}{3}\mu$.  These models are spatially homogeneous but not spatially
 isotropic.  The manifold $\M$ for these models is $\R^4$ and they
have a metric $\vec{g}$, in comoving coordinates, of the form
\be
ds^2= -\:  Adt^2 +t\left[ A^{-1}dx^2+A^2 b^{-2}(dy^2+f^2dz^2) \right],
                \quad t>0 \, ,
\end{equation}
\bea
\text{where}& &A=1-\frac{4\epsilon b^2t}{9} \; , \; \;
\quad b\text{ is a constant},
                \nonumber \\
& &f(y)= \left\{ \begin{array}{ll}
                \sin y & \quad\epsilon = +1 \\
                \sinh y & \quad\epsilon = -1
                \end{array} \right.\nonumber,
\eea
and the fluid flow $\vec{u}$ is
\be
\vec{u} = A^{-\frac{1}{2}} \, \frac{\partial}{\partial t} \: .
\end{equation}
Now define a cosmic time function $T$ by
\be
T(t) = \sqrt{2t} \, .
\end{equation}
Let $\s\M$ be the manifold $\R^4$ covered by the coordinate patch $(T,x,y,z)$,
 where $T\in \R$.  Thus $\M$ is the open submanifold $T>0$ of $\s\M$, thereby
 satisfying requirement (1) of the definition of an IS.

Now define a new coordinate $\overline{x}$ by
$\overline{x}=\frac{1}{\sqrt{2}}x$,
and a new constant $\overline{b}$ by
$\overline{b} = \sqrt{2}\thinspace b$.  Hence
\be
dt = T \, dT \; \quad\text{and}\quad \; d\overline{x} =
\frac{1}{\sqrt{2}}dx \: .
\end{equation}


\noindent
Rewriting the metric $\vec{g}$ in terms of the new $T$ and $\overline{x}$
coordinates, we obtain
\be
ds^2 = \, T^2 \left( - \: AdT^2 +A^{-1}d{\overline{x}}^2
                +A^2(\overline{b})^{-2}(dy^2 +f^2dz^2)  \right), 
\end{equation}
\be
\text{and}\quad\; \vec{u} \, = \, A^{-\frac{1}{2}}T^{-1}
\: \frac{\partial}{\partial T} \: ,
                \quad \quad \text{where}
                \quad\; A=1-\frac{\epsilon \overline{b}^2T^2}{9} \: .
\end{equation}
Now define the conformal factor $\Omega$ by
\be
\Omega(T)= T \, , \quad\text{where }T\geq0 \: .
\end{equation}
Clearly this conformal factor satisfies requirements (3) and (4) of the
definition of an IS, with $\lambda=0$.

The conformally related metric $\s \vec{g}$ is then given by
\be
{\s ds}^2 =  - \: A dT^2 +A^{-1}d{\overline{x}}^2
                +A^2(\overline b)^{-2}(dy^2 +f^2dz^2) \: ,
\end{equation}
and is certainly $C^3$ and non-degenerate on an open neighbourhood of $T=0$,
 thereby satisfying requirement (2) of the definition of an IS.
Thus, the Kantowski-Sachs models do indeed admit an isotropic
singularity---this was first shown by Goode and Wainwright.\cite{GW85}

From Definition (\ref{defn:fluid_congruence}), the unphysical fluid flow
$\s\vec{u}$ is given by
\be
\s\vec{u} \, =  \, A^{-\frac{1}{2}} \: \frac{\partial}{\partial T} \: .
\end{equation}
It is clear that $\s\vec{u}$ is $C^3$ on an open neighbourhood of $T=0$ in
 $\s\M$, and hence $\vec{u}$ is regular at the isotropic singularity.

The Weyl tensor ${C^a}_{bcd}$ is bounded at the IS, with some components
having a non-zero limit.  Splitting the Weyl tensor up into its electric,
 $E_{ab}$, and magnetic, $H_{ab}$, parts, it is found that the magnetic part
 is everywhere zero and the electric part is bounded at the IS, with some
components having a non-zero limit.


\section{Szekeres models}
The Szekeres models\cite{Szekeres75} are irrotational, geodesic, pressure-free
 dust solutions.  In comoving
coordinates, the metric $\vec{g}$ and fluid congruence $\vec{u}$ for these
models are given by
\be
ds^2=- \: dt^2+ \, t^{\frac{4}{3}}(dx^2+dy^2+Z^2dz^2) \, , \quad t>0 \, ,
        \qquad\vec{u}=\frac{\partial}{\partial t} \, ,
\end{equation}
where
\be
Z=A+k_+t^{\frac{2}{3}}+k_{-}t^{-1} \, ,
\quad \quad A=ax+by+c+\frac{5}{9}k_+(x^2+y^2) \, ,
\end{equation}


\noindent
and $a, b, c, k_+, k_-$ are arbitrary smooth functions of z.

We will only examine the Szekeres models in which the decaying mode is absent,
 i.e., when $k_-=0$.  This subclass has previously been shown by Goode and
Wainwright\cite{GW82} to have a Friedmann-like singularity and was first
explicitly shown to have an IS in a paper by Goode, Coley, and
Wainwright.\cite{GCW92}  The manifold $\M$ for these models is $\R^4$.
Now define a cosmic time function $T$ by
\be
T(t) = 3 \, t^{\frac{1}{3}},\quad\text{and hence},
\quad dt \, = \, \frac{T^2}{9} \, dT \: .
\end{equation}
Let $\s\M$ be the manifold $\R^4$ covered by the coordinate patch $(T,x,y,z)$,
 where $T\in \R$.  Thus $\M$ is the open submanifold $T>0$ of $\s\M$, thereby
 satisfying requirement (1) of the definition of an IS.

Rewriting the metric $\vec{g}$ in $(T,x,y,z)$ coordinates, we obtain
\be
ds^2 \, = \, \frac{T^4}{81} \: (- \: dT^2+dx^2+dy^2+Z^2dz^2) \: ,
\end{equation}
where
\be
Z \, = \, A+k_+ \frac{T^2}{9} \: .
\end{equation}
Now define the conformal factor $\Omega$ by
\be
\Omega(T)=\frac{T^2}{9} \, ,\quad\text{where  }T\geq 0 \: .
\end{equation}
Clearly this conformal factor satisfies requirements (3) and (4) of the
definition of an IS, with $\lambda=\smallfrac{1}{2}$.

The conformally related metric $\s \vec{g}$ and fluid flow $\s\vec{u}$ are
given
by
\be
\s ds^2=- \: dT^2+dx^2+dy^2+Z^2dz^2 \, ,
\qquad\s\vec{u}=\frac{\partial}{\partial T} \: .
\end{equation}
It is readily seen that condition (2) of the definition of an IS is satisfied
 and that the fluid flow $\vec{u}$ is regular at the IS. Thus the Szekeres
models with $k_-=0$ certainly admit an IS.

The Weyl tensor behaves qualitatively the same as the Kantowski-Sachs Weyl
tensor. It is bounded at the IS, with some components having a non-zero
limit, the magnetic part is everywhere zero, and the electric part is bounded
 at the IS, with some components having a non-zero limit.



\section{Mars models}
Mars\cite{Mars95} has found three types of solution corresponding to
a perfect fluid with an Abelian two-dimensional group of isometries
 acting orthogonally transitively on spacelike 2-surfaces and such that both
Killing vectors are integrable.  In comoving coordinates, the third type of
solution has a metric $\vec{g}$ of the form
\bea
ds^2 \: = \: -\;\frac{e^{at+\epsilon c^2 e^{2a(t+x)}}}
          {1\plus \, \epsilon e^{-2at}\plus \, \beta e^{-6at}} \: dt^2
                        &\!\!\!\!+\!\!\!\!&e^{at+\epsilon c^2 e^{2a(t+x)}}dx^2 
				\nonumber \\
                +\;e^{a(t-x)+2ce^{ax}}dy^2 
				&\!\!\!\!+\!\!\!\!& e^{a(t-x)-2ce^{ax}}dz^2\:,
\eea
where $a,c,\epsilon,\beta$ are constants with $a>0,\epsilon = +1,\beta\geq 0$.
The fluid flow  $\vec{u}$ is given by
\be
\vec{u} \, = \, \sqrt{\frac{1+\epsilon e^{-2at}+\beta e^{-6at}}
      {e^{at+\epsilon c^2 e^{2a(t+x)}}}} \; \frac{\partial}{\partial t} \; .
\end{equation}
The manifold $\M$ for these models is $\R^4$, and it is to be noted that they
 all have a big-bang singularity at $t=-\infty$.  We can define a cosmic time
 function $T$ for these models by
\be
 T=e^{at}, \quad\text{and hence,} \quad dt =\frac{1}{aT} \: dT \: .
\end{equation}
Let $\s\M$ be the manifold $\R^4$ covered by the coordinate patch $(T,x,y,z)$,
 where $T\in \R$.  Thus $\M$ is the open submanifold $T>0$ of $\s\M$, thereby
 satisfying requirement (1) of the definition of an IS.

Rewriting the metric $\vec{g}$ in $(T,x,y,z)$ coordinates, we obtain
\bea
& &ds^2= \: \minus \: T\left( \frac{e^{\epsilon c^2 T^2e^{2ax}}}
         {(1\plus \, \epsilon T^{-2}\plus \, \beta T^{-6})a^2T^2} \: dT^2
                      + e^{\epsilon c^2 T^2e^{2ax}}dx^2 \right.\nonumber \\
 & &\quad\qquad\qquad\qquad\left.+ \: e^{-ax+2ce^{ax}}dy^2+ e^{-ax-2ce^{ax}}dz^2
                \mbox{\rule{0mm}{6mm}}\right) ,
\eea
and the fluid flow $\vec{u}$ becomes
\be
\vec{u} \, = \, aT \: \sqrt{\frac{1+\epsilon T^{-2}+\beta T^{-6}}
                        {Te^{\epsilon c^2 T^2e^{2ax}}}}
                        \; \frac{\partial}{\partial T} \; .
\end{equation}

For $\beta > 0$, there is a limiting $\gamma$-law equation of state,
$p=(\gamma\minus1)\mu$, for the perfect fluid, with
$\gamma = \frac{14}{3}$.  If $\beta = 0$, then the fluid is a stiff fluid,
 i.e., it has an exact $\gamma$-law equation of state with $\gamma=2$.


The vorticity scalar, shear scalar, and acceleration for the fluid flow
$\vec{u}$ are given by
\bea
\omega^2 &= &0 \, , \\
\sigma^2 &=& \frac{a^2\epsilon^2 c^4}{3}
        \left[ T^3 \plus \, \epsilon T \plus \, \frac{\beta}{T^3} \right]
        e^{-\epsilon c^2T^2e^{2ax}+4ax} \, , \\
\dot{u}^b &=& a\epsilon c^2 T^2e^{2ax} \delta^b_1 \,.
\eea
It follows that the Mars models are irrotational, perfect fluid models which
have \emph{ non-geodesic} fluid flows.  Now restricting ourselves to the
subclass of these models which satisfy  $\beta=0$, we can define the
conformal factor $\Omega$ by
\be
\Omega(T)=\sqrt{T} \, ,\quad T>0 \, ,\qquad \Omega(0)=0 \, .
\end{equation}
Clearly this conformal factor satisfies requirements (3) and (4) of the
definition of an IS, with $\lambda=-1$.

The conformally related metric $\s \vec{g}$ and fluid flow $\s \vec{u}$ are
given by
\bea
\s ds^2  &=& \: \minus \; 
		\frac{e^{\epsilon c^2 T^2e^{2ax}}}
                        {a^2(\epsilon\plus T^2)} \; dT^2
                 + \, e^{\epsilon c^2 T^2e^{2ax}}dx^2 \nonumber \\
                        & &\qquad\qquad\qquad
              + \: e^{-ax+2ce^{ax}}dy^2+ \, e^{-ax-2ce^{ax}}dz^2 \: ,  \\
\text{and } \quad\s\vec{u} \: &=& \: a \; 
					\sqrt{\frac{\epsilon+T^2}
                        {e^{\epsilon c^2T^2 e^{2ax}}}}
                               \: \frac{\partial}{\partial T} \: .
\eea

It is readily seen that condition (2) of the definition of an IS is satisfied,
 and that the fluid flow $\vec{u}$ is regular at the IS as well as being
orthogonal to the IS.  Thus the Mars models with $\beta=0$ do admit an IS.
This fact was stated by Mars\cite{Mars95} but not shown explicitly until this
paper.

The Weyl tensor ${C^a}_{bcd}$ is bounded at the IS, with some components
having a non-zero limit.  Splitting the Weyl tensor up into its electric,
 $E_{ab}$, and magnetic, $H_{ab}$, parts, it is found that the magnetic part
 limits to zero, as the IS is approached, and the electric part is bounded
at the IS, with some components having a non-zero limit.


\section{Other models}

In this section, we list all other cosmological models which are known to
admit an IS. For some models, the authors present the actual conformal
structure which yields the IS, whilst for the others, the authors simply
claim that the models admit an IS.


\subsection{Collins 71 models - Bianchi type VI${}_h$ }
These models, found by Collins\cite{Collins71} in 1971, are spatially
 homogeneous, irrotational, geodesic, perfect fluid cosmological models with a
 $\gamma$-law equation of state, $p=(\gamma-1)\mu$.  A subclass of these models
 was first shown to admit an IS in 1984 by Wainwright and
Anderson.\cite{WA84,Tod92} For this subclass, the fluid flow has non-zero
shear, although the shear vanishes at the IS. The magnetic part of the Weyl
tensor limits to zero as the IS is approached, while the electric part of the
Weyl tensor is bounded at the IS, with some components having a non-zero limit.

\subsection{Bondi models}
The spherically symmetric dust models of Bondi\cite{Bondi47} are irrotational
and geodesic, and the magnetic part of the Weyl tensor is everywhere zero.  A
subclass of these models was shown to admit an IS in 1992 by Tod.\cite{Tod92}
For this subclass, the fluid flow has non-zero shear, with the shear vanishing
 at the IS, and the electric part of the Weyl tensor is bounded at the IS,
with some components having a non-zero limit.


\subsection{Others}

Goode, Coley, and Wainwright\cite{GCW92} state that Wainwright and
Hsu\cite{WH89} have shown that all Bianchi classes contain solutions which
admit an IS, although the only known non-FRW solutions are those of
Wainwright and Anderson.\cite{WA84}

The plane-symmetric, stiff matter solutions of Tabensky and Taub\cite{TT73}
are  irrotational and geodesic, and the magnetic part of the Weyl tensor is
everywhere zero.  Tod\cite{Tod87} states that a certain subclass of these
solutions (those with the coefficient $b$ set to zero) admit an IS.
Rendall\cite{Rendall95} has actually shown that, for a specific case, the
 Tabensky-Taub solutions admit an IS.

Mimoso and Crawford\cite{MC93} have found a class of spatially homogeneous,
irrotational, shear-free, geodesic cosmological models, with an imperfect fluid
matter source, which they claim admits an IS. The matter source is, in fact, an
anisotropic fluid without heat flux, and the magnetic part of the Weyl tensor
is everywhere zero.


\section{Discussion}

In Sections $2,3,4,$ and $5$, we have listed, and examined, in varying detail,
all cosmological models which are known to admit an IS. We note that all
models discussed have a perfect fluid source, except for the Mimoso-Crawford

models, which have an imperfect fluid source corresponding to an
anisotropic fluid without heat flux. The conformal structure which yields
an IS for the Mimoso-Crawford models has yet to be presented.

It is instructive, at this point, to place all the perfect fluid models
in a table, categorised according to their physical attributes---the
fluid vorticity, shear, and acceleration; the Weyl tensor; the electric
and magnetic parts of the Weyl tensor; the equation of state for the
perfect fluid.

\begin{table}[ht]
\small
\centering
\begin{tabular}{|l|c|c|c|c|c|c|l|}\hline
Models          &$\omega_{ab}$  &$\sigma_{ab}$  &$\dot{u}^a$
                        &${C^a}_{bcd}$  &$E_{ab}$&$H_{ab}$ &$p=p(\mu)$\\\hline
FRW             & 0             & 0             & 0
                        & 0             & 0      & 0      &yes
                                        (a subclass)\\ \hline
Kantowski-Sachs & 0             &(a)            & 0
                        &(b)            &(b)     & 0      &yes
                                        ($p=\smallfrac{1}{3}\mu$)\\\hline
Szekeres        & 0             &(a)            & 0
                        &(b)            &(b)     & 0      &yes
                                        ($p=0$)\\\hline
Bondi           & 0             &(a)            & 0
                        &(b)            &(b)     & 0      &yes
                                        ($p=0$)\\ \hline
Tabensky-Taub   & 0             &(a)            & 0
                        &(b)            &(b)     & 0      &yes
                                        ($p=\mu$)\\\hline
Collins 71 & & & & & & &\\
Bianchi type VI${}_h$& 0        &(a)            & 0
                        &(b)            &(b)     & (a)    &yes
                                        ($p=(\gamma\minus 1)\mu$)\\ \hline
Mars 95         & 0             &(a)            & (a)
                        &(b)            &(b)     & (a)    &yes
                                        ($p=\mu$)\\\hline
\end{tabular}
\caption{Perfect fluid cosmological models with an IS: (a) means that the
 relevant tensor is non-zero away from the IS but vanishes as the IS is
approached, (b) means that the relevant tensor components are bounded as
the IS is approached, with some components having a non-zero limit.}
\end{table}

A number of interesting characteristics are apparent from the table. Apart from
the FRW models, all other perfect fluid models which are known to admit an IS
have an exact $\gamma$-law equation of state. This poses the
question: \emph{ do there exist any non-FRW barotropic perfect fluid
cosmological models which admit an IS, yet which do not have an
exact $\gamma$-law equation of state?}

All perfect fluid models in the table are irrotational. Indeed, the
General Vorticity Result of Scott\cite{SSGVR} proves that \emph{ a
barotropic perfect fluid cosmological model with non-zero vorticity cannot
admit an IS.} We also note that there is one class of models in the table,
namely the Mars models, which have a non-geodesic fluid flow. We deduce from
this that, unlike irrotationality, geodesicity of the fluid flow is not a
necessary condition for a barotropic perfect fluid cosmological model to
satisy in order to admit an IS.

It remains an important and open problem to produce a precise set of
necessary and sufficient conditions which a barotropic perfect fluid
cosmological model must satisfy in order to admit an IS at which the
fluid flow is regular.

We also note from the table that the FRW models are the only models
present with Weyl tensor components ${C^a}_{bcd}$ which all vanish as the
IS is approached. This lends weight to what is known as the FRW
conjecture\cite{GW85,Tod87,Scott88b}.


\begin{conj}[FRW]
If a space-time is
\begin{enumerate}
\item a $C^3$ solution of the Einstein field equations with a barotropic
        perfect fluid source, and
\item the unit timelike fluid congruence is regular at an isotropic
        singularity, and
\item the Weyl Curvature Hypothesis holds,
\end{enumerate}
then the space-time is necessarily a Friedmann-Robertson-Walker model.
\end{conj}


\section*{References}

\newcommand{\journal}[7]{#1, \textit{#3}\textbf{#4}, #6 (#5)}

\newcommand{\proc}[8]{#1,  \textit{ #3}  ed. #4 (#5) (#6) pp#7-#8}

\newcommand{\phd}[3]{#1, \textit{ Ph.D. thesis} #2 (#3)}

\newcommand{\AOP}{A.P.\ }
\newcommand{\CQG}{Class.\ Quantum Grav.\ }
\newcommand{\GRG}{Gen.\ Rel.\ Grav.\ }
\newcommand{\PR}{Phys.\ Rev.\ }
\newcommand{\CMP}{Commun.\ math.\ Phys.\ }
\newcommand{\MNRAS}{Mon.\ Not.\ R.\ astr.\ Soc.\ }
\newcommand{\PL}{Phys.\ Lett.\ }


\end{document}